\documentclass[11pt]{article}
\newcommand{\Comment}[1]{{}}
\usepackage{amsfonts,amsthm,amsmath,amssymb,slashed}
\usepackage[textwidth = 430 pt, textheight = 630 pt]{geometry}

\Comment{\usepackage{color}
\definecolor{MyDarkBlue}{rgb}{0.15,0.15,0.45}
\usepackage[linktocpage=true]{hyperref}
\hypersetup{
colorlinks=true,
citecolor=MyDarkBlue,
linkcolor=MyDarkBlue,
urlcolor=MyDarkBlue,
pdfauthor={Kurt Hinterbichler, Justin Khoury and Horatiu Nastase},
pdftitle={Towards a UV completion for chameleon scalar theories},
pdfsubject={hep-th}
}

\usepackage[numbers,sort&compress]{natbib}
\usepackage{hypernat}}
\usepackage{graphicx}

\newcommand\ignore[1]{}
\def\one{{\,\hbox{1\kern-.8mm l}}}

\def\a{\alpha}\def\b{\beta}

\def\d{\partial}

\newcommand{\Cset}{{\,\,{{{^{_{\pmb{\mid}}}}\kern-.45em{\mathrm C}}}}}

\newcommand{\be}{\begin{equation}}
\newcommand{\bea}{\begin{eqnarray}}

\newcommand{\ee}{\end{equation}}
\newcommand{\eea}{\end{eqnarray}}

\newcommand{\half}{\frac{1}{2}}

\providecommand{\lsim}{\lesssim}
\providecommand{\gsim}{\gtrsim}

\begin{document}

\renewcommand{\thefootnote}{\fnsymbol{footnote}}

\makeatletter
\@addtoreset{equation}{section}
\makeatother
\renewcommand{\theequation}{\thesection.\arabic{equation}}

\rightline{}
\rightline{}
   \vspace{1.8truecm}

%\begin{flushright}
% preprint nrs.
%\end{flushright}

\vspace{10pt}

%%%%%%%%%%%%%%%%%

\begin{center}
{\LARGE \bf{\sc Towards a UV Completion of Chameleons in String Theory}}
\end{center} 
 \vspace{1truecm}
\thispagestyle{empty} \centerline{
{\large \bf {\sc Kurt Hinterbichler${}^{a,}$}}\footnote{E-mail address: \Comment{\href{mailto:kurthi@physics.upenn.edu}}{\tt kurthi@physics.upenn.edu}},
{\large \bf {\sc Justin Khoury${}^{a,}$}}\footnote{E-mail address: \Comment{\href{mailto:jkhoury@sas.upenn.edu}}{\tt jkhoury@sas.upenn.edu}}
{\bf{\sc and}}
    {\large \bf {\sc Horatiu Nastase${}^{b,}$}}\footnote{E-mail address: \Comment{\href{mailto:nastase@ift.unesp.br}}{\tt 
    nastase@ift.unesp.br}} 
                                                           }

\vspace{1cm}

\centerline{{\it ${}^a$ 
Center for Particle Cosmology, Department of Physics and Astronomy,}}
 \centerline{{\it University of Pennsylvania, 209 South 33rd Street, }} \centerline{{\it Philadelphia, PA 19104, USA}}

\vspace{.8cm}
\centerline{{\it ${}^b$ 
Instituto de F\'{i}sica Te\'{o}rica, UNESP-Universidade Estadual Paulista}} \centerline{{\it 
R. Dr. Bento T. Ferraz 271, Bl. II, Sao Paulo 01140-070, SP, Brazil}}

\vspace{2truecm}

%%%%%%%%%%%%%%%%%
\thispagestyle{empty}

\centerline{\sc Abstract}

\vspace{.4truecm}

\begin{center}
\begin{minipage}[c]{380pt}
{\noindent Chameleons are scalar fields that couple directly to ordinary matter with gravitational strength, but which nevertheless evade
the stringent constraints on tests of gravity because of properties they acquire in the presence of high ambient matter density. 
Chameleon theories were originally constructed in a bottom-up, phenomenological fashion, with potentials and matter couplings designed to hide
the scalar from experiments. In this paper, we attempt to embed the chameleon scenario within string compactifications, 
thus UV completing the scenario. We look for stabilized potentials that can realize a screening mechanism, and we find that the 
volume modulus rather generically works as a chameleon, and in fact the supersymmetric potential
used by Kachru, Kallosh, Linde and Trivedi (KKLT) is an example of this type. We consider all constraints from tests of gravity,
allowing us to put experimental constraints on the KKLT parameters.}
\end{minipage}
\end{center}

\vspace{.5cm}

\setcounter{page}{0}
\setcounter{tocdepth}{2}

\newpage

%\tableofcontents
\renewcommand{\thefootnote}{\arabic{footnote}}
\setcounter{footnote}{0}

\linespread{1.1}
\parskip 4pt

%{}~
%{}~

\section{Introduction}
\ \ \ \ \
A generic prediction of string theories and other higher dimensional attempts at unification is the existence of light, gravitationally coupled scalars.  For instance, string compactifications generally predict the existence of many such scalars, which can be thought of geometrically as parametrizing moduli spaces, that is, spaces of choices in the compactification ingredients (for example the size and shape of the compactification manifold).  These scalars can be put to good use on large scales, for instance they may play a crucial role in explaining dark energy as quintessence fields \cite{Ratra:1987rm,Caldwell:1997ii}, and they are generically used in infrared-modified gravity theories~\cite{Dvali:2000hr,ArkaniHamed:2003uy,Khoury:2003aq,Khoury:2003rn,Gubser:2004uf,deRham:2007xp,deRham:2007rw,Nicolis:2008in,Sotiriou:2008rp,DeFelice:2010aj}. See~\cite{Jain:2010ka} for a recent review.

But there is the perplexing fact that no sign of such a fundamental scalar field has ever been seen experimentally, despite many searches and tests.  These searches and tests are designed to detect local effects in the laboratory and in the solar system from the fifth forces that would naively be expected if such scalars existed~\cite{Fischbach:1999bc,Will:2005va}. 

It is possible, however, that cosmologically interesting scalars are present in nature, but have managed to avoid detection in all such experiments.  Three broad classes of theoretical mechanisms have been developed to explain why such light scalars may not be visible to experiments performed near the Earth.  See~\cite{Khoury:2010xi} for a review.  Experiments generally look for canonical linear scalars, whereas these mechanisms all rely on the presence of non-linearities in the scalar field action.  

The first such mechanism, the chameleon mechanism~\cite{Khoury:2003aq,Khoury:2003rn,Gubser:2004uf,Upadhye:2006vi,Brax:2004qh,Brax:2004px,Mota:2006ed,Mota:2006fz,Brax:2007hi,Brax:2007vm,Brax:2009bk,Burrage:2007ew,Burrage:2008ii,Steffen:2009sc,Upadhye:2009iv}, essentially makes the effective mass of the scalar depend on the local density of ordinary matter. Deep in space, where the local mass density is low, the scalars would be light and would display their effects, but near the Earth, where experiments are performed, and where the local mass density is high, they would acquire a mass, making their effects short range and unobservable.   The chameleon mechanism requires non-minimal coupling to matter and a suitable form for the scalar potential.

The second such mechanism, the Vainshtein mechanism~\cite{Vainshtein:1972sx,ArkaniHamed:2002sp,Deffayet:2001uk}, operates when the scalar has derivative self-couplings which become important near matter sources such as the Earth. The strong coupling near sources boosts the kinetic terms, which means, after canonical normalization, that the couplings to matter are weakened. Thus the scalar decouples itself and becomes invisible to experiments. This mechanism is central to the phenomenological viability of massive gravity~\cite{Vainshtein:1972sx,ArkaniHamed:2002sp,Deffayet:2001uk,Gabadadze:2009ja,deRham:2009rm,deRham:2010gu,deRham:2010ik,deRham:2010tw,deRham:2010kj}, degravitation theories~\cite{Dvali:2002pe,ArkaniHamed:2002fu,Dvali:2007kt}, brane-world modifications of gravity~\cite{Dvali:2000hr,deRham:2007xp,deRham:2007rw,deRham:2009wb,deRham:2010rw,Agarwal:2009gy} and galileon scalar theories~\cite{Nicolis:2008in,Deffayet:2009wt,Deffayet:2009mn,Chow:2009fm,Silva:2009km,DeFelice:2010pv,Mota:2010bs,Hinterbichler:2010xn,Andrews:2010km,Goon:2010xh,Padilla:2010de,Padilla:2010tj,Padilla:2010ir}.

The third such mechanism, the symmetron~\cite{Hinterbichler:2010es} (see also~\cite{Olive:2007aj,Pietroni:2005pv}), works by exploiting spontaneous symmetry breaking.  It requires a symmetry-breaking, Mexican-hat type potential and non-minimal couplings to matter.  The non-minimal couplings to matter are such that the vacuum expectation value (VEV) of the scalar depends on the local mass density, becoming large in regions of low mass density, and small in regions of high mass density.   The same non-minimal couplings cause the couplings of the scalar fluctuations to matter to be proportional to the VEV, so that the scalar couples with gravitational strength in regions of low density, but is decoupled and screened in regions of high density.

The chameleon and symmetron mechanisms were constructed in a bottom up, phenomenological fashion, with potentials and matter couplings rigged up to hide the scalar from experiments.  The chameleon has been criticized for being unnatural~\cite{Jain:2010ka}, and while the symmetron has a nicer-looking potential, they are both effective theories and must eventually find an embedding in some larger well-motivated theory.  (The Vainshtein mechanism was originally discovered in the context of massive gravity, also an effective theory that requires UV completion \cite{ArkaniHamed:2002sp}.  In general, any Vainstein mechanism requires a UV completion because it relies on derivatively coupled interactions, which are non-renormalizable).

Here we attempt to find a chameleon or symmetron type mechanism within a string compactification, thus UV completing the scenario (we do not 
consider here embedding the Vainstein mechanism).  We will attempt to use the volume modulus as the chameleon (for an attempt to use the 
dilaton, see~\cite{Brax:2010gi}).  The moduli in string compactifications should eventually be stabilized, and in particular the 
volume modulus governing the KK scale of the extra dimensions, and a potential will be generated for 
these scalars.  We look for stabilized potentials that can realize a screening mechanism, focusing on a rather general class of such potentials, 
and we find that the potential used by Kachru, Kallosh, Linde and Trivedi (KKLT)~\cite{Kachru:2003aw} is in fact an example of this type.
\footnote{For an attempt to use an RS-type radion in a supersymmetric model as a chameleon with quintessence, see \cite{Brax:2004ym}, for an 
analysis of Kahler moduli within general ${\cal N}=1$ supergravity as inverse power law-type chamelons see \cite{Brax:2006np}, and for some high
density effects in the usual KKLT scenario see \cite{Conlon:2010jq}.}

The KKLT scenario involves an anti-D3-brane at the bottom of a compactified Klebanov-Strassler solution~\cite{Klebanov:2000hb}. The presence of the anti-D3-brane breaks 
supersymmetry in a controllable way, but one can consider the model without the supersymmetry breaking term as well. Then the superpotential for the complexified volume 
modulus $\varrho$ contains a tree-level constant piece, no perturbative corrections, and a nonperturbative exponential piece $\sim e^{ia\varrho}$ coming from 
two possible effects: Euclidean D3-brane instantons and gluino condensation on D7-branes. The other moduli are considered to be stabilized. Here $a>0$ for the two nonperturbative effects considered, but cases with $a<0$ have been considered before, even in the context of KKLT~\cite{Abe:2005rx}, as well as in other cases, {\it e.g.}~\cite{Quevedo:1996sv,Burgess:1997pj}. It will turn out that for our chameleon model we will need $a<0$.

We begin in  Sec.~\ref{generic} by reviewing the salient features of chameleon field theories, focusing on the thin-shell screening mechanism central to their phenomenological viability.
In Sec.~\ref{chamdimred}, we show that the volume modulus of generic dimensional reduction is a suitable candidate to exhibit chameleon properties, and we devise a useful phenomenological potential which captures its general features.  In Sec.~\ref{egkklt}, we argue that the potential introduced by
KKLT~\cite{Kachru:2003aw} is in fact of the desired form (in the case where $a<0$ already mentioned above). We consider all constraints from
tests of gravity on the general, phenomenological potential in Sec.~\ref{consastro}, and specialize them to the KKLT potential in Sec.~\ref{KKLTapply}, allowing us to put experimental constraints on the KKLT parameters. We recap the key results and discuss future directions in Sec.~\ref{conclude}.

\section{Generic Chameleon Potentials}\label{generic}
\ \ \ \ \
We focus on chameleon and symmetron type models, both of which are encompassed by the following generic scalar tensor action~\cite{Khoury:2003aq} (with mostly-plus metric signature)
\be
S=\int {\rm d}^4x\sqrt{-g}\left[ {M_{\rm Pl}^2\over 2} R[g]-\half g^{\mu\nu}\partial_\mu \phi\partial_\nu\phi-V(\phi)\right]
+ \int {\rm d}^4x\; {\cal L}_{\rm m}[\tilde{g},\psi]\,,
\label{symmetronlagrangian}
\ee
where $M_{\rm Pl} \equiv (8\pi G_{\rm N})^{-1}$ is the (reduced) Planck scale.
The matter fields $\psi$, described by ${\cal L}_{\rm m}$, are universally coupled to the metric $\tilde{g}_{\mu\nu}$,
which is conformally related to the Einstein frame metric $g_{\mu\nu}$ by
\be 
\tilde{g}_{\mu\nu} =  A^2(\phi) g_{\mu\nu}\,.
\label{conf}
\ee
The scalar field equation is
\be
\square\phi-V_{,\phi}+A^3(\phi)A_{,\phi} \tilde{T}=0\,,
\ee
where $\tilde{T}= \tilde{T}_{\mu\nu}\tilde{g}^{\mu\nu}$ is the trace of the Jordan frame energy momentum tensor,
$\tilde{T}_{\mu\nu}= -(2/\sqrt{-\tilde{g}})\delta{\cal L}_{\rm m}/ \delta \tilde{g}^{\mu\nu}$, which is covariantly conserved by the Jordan frame covariant derivative: 
$\tilde{\nabla}_\mu  \tilde{T}^{\mu}_{\ \nu}=0$.

We will be interested in solar system and galactic scenarios, so we may ignore the non-linearities in gravity and the back-reaction of the scalar field on the metric.  This allows us to treat the (non-linear) scalar on its own, in a fixed Minkowski background.  For astrophysical objects, we may also use the idealization of a spherically symmetric pressureless source.  Written in terms of the density $\rho=A^{3}\tilde\rho$, which is conserved in Einstein frame, the scalar field equation takes the form
\be 
{{\rm d}^2\phi\over {\rm d}r^2}+{2\over r}{{\rm d}\phi\over {\rm d}r} =V_{,\phi}+ A_{,\phi}\rho\,.
\label{sphericalequation} 
\ee
For cases of roughly homogeneous $\rho$, such as the interior or exterior of a star or galaxy, the field thus evolves according to an effective potential
\be 
V_{\rm eff}(\phi)=V(\phi)+\rho A(\phi)\,.
\label{Veff}
\ee
The effective potential thus depends linearly on the local matter density, and this is the key to the screening mechanisms.

In the chameleon case, by suitably choosing $V(\phi)$ and $A(\phi)$ the effective potential can develop a minimum  at some finite field value $\phi_{\rm min}$ in the presence of background matter density, where the mass of the chameleon field,
\be
m_{\rm min}^2 = V_{,\phi\phi}(\phi_{\rm min}) + A_{,\phi\phi}(\phi_{\rm min})\rho\,,
\ee
is sufficiently large to evade local constraints. Assuming $A(\phi)$ is monotonically increasing, for concreteness, the general conditions that $V$ must satisfy are~\cite{Khoury:2003aq,Khoury:2003rn,Brax:2004qh,Brax:2004px}: $i)$ $V_{,\phi} < 0$ over the relevant field range, in order to balance the potential against the density term; $ii)$ since $V_{,\phi\phi}$ typically gives the dominant contribution to $m_{\rm min}$, stability requires $V_{,\phi\phi}> 0$; $iii)$ in order for $m_{\rm min}$ to increase with $\rho$, we demand that $V_{,\phi\phi\phi} < 0$.

A prototypical potential satisfying these conditions is the inverse power-law form, $V(\phi) = M^{4+n}/\phi^n$, where $M$ is some mass scale. 
For the coupling function, a generic form that makes contact with Brans-Dicke theories is 
\be
A(\phi) = e^{g\phi/M_{\rm Pl}}\approx 1 + g\phi/M_{\rm Pl}\,,
\label{confbis}
\ee
where we have used the fact that $\phi\ll M_{\rm Pl}$ over the relevant field range. The parameter $g$ is implicitly assumed to be~${\cal O}(1)$, corresponding to gravitational strength coupling. 
For $g > 0$, the effective potential, $V_{\rm eff}(\phi) = M^{4+n}\phi^{-n} + \rho e^{g\phi/M_{\rm Pl}}$, displays a minimum at $\phi_{\rm min} \sim \rho^{-1/(n+1)}$. It follows that the mass of small fluctuations around the minimum, $m^2_{\rm min}\sim \rho^{(n+2)/(n+1)}$, is an increasing function of the background density, as desired. In this paper, we will consider potentials of a different form, suggested by 
flux compactifications in string theory. 

The density-dependent mass results in a further decoupling effect outside sufficiently massive objects, due to the so-called {\it thin-shell} effect. Consider a spherical source of
radius ${\cal R}$ and density $\rho_{\rm in}$ (with corresponding effective minimum at $\phi_{\rm min-in}$) embedded in a homogeneous medium of density $\rho_{\rm out}$ (with corresponding $\phi_{\rm min-out}$). If the source is sufficiently massive, the scalar field at the core of the source is oblivious to the exterior matter and is therefore pinned near
$\phi_{\rm min-in}$ in the core.  As we move up in radius towards the surface, the scalar must eventually reach the asymptotic value $\phi_{\rm min-out}$ far away, so
it must deviate substantially from $\phi_{\rm min-in}$ near the surface. Consequently, inside the object the evolution of $\phi$ is essentially confined within
a thin-shell $\Delta {\cal R} \equiv {\cal R}- {\cal R}_{\rm roll}$, where $ {\cal R}_{\rm roll}$ is the radius at which the field starts to deviate from its value at the core.  

Thus if a spherical object is sufficiently large and/or massive, in a way that will be made precise shortly, then 
\be
\phi\simeq \phi_{\rm min-in}\qquad  {\rm for}\;\;\;  0<r<{\cal R}_{\rm roll}\,.
\ee
Within the shell, meanwhile, the field evolution is dominated by the $\rho$ term, hence
\be
\phi\simeq \phi_{\rm min-in} + \frac{g}{2M_{\rm Pl}} \rho_{\rm in}(r-{\cal R}_{\rm roll})^2\qquad {\rm for}\;\;\; {\cal R}_{\rm roll}<r<{\cal R}\,.
\ee
Matching this to the exterior solution at $r = {\cal R}$ allows one to solve for ${\cal R}_{\rm roll}$. Explicit computation shows that~\cite{Khoury:2003aq,Khoury:2003rn}
\be
\frac{\Delta {\cal R}}{{\cal R}} \equiv \frac{{\cal R}-{\cal R}_{\rm roll}}{{\cal R}} = \frac{\phi_{\rm min-out} - \phi_{\rm min-in}}{6g M_{\rm Pl}\Phi}\,,
\label{thin}
\ee
where $\Phi\equiv \rho_{\rm in}{\cal R}^2/6M_{\rm Pl}^2$ is the surface Newtonian potential. In other words, the shell thickness is determined by the difference in $\phi$ values
relative to the difference in gravitational potential between the surface and infinity. Note, however, that the thin shell can never be thinner than the
Compton wavelength of the field inside the object. In other words, $\Delta {\cal R}/{\cal R}$ is given by the largest of $m^{-1}_{\rm min-in}{\cal R}^{-1}$ and
the right hand side of~(\ref{thin}).

Since field gradients are essentially confined to the thin shell, the exterior profile is suppressed by the thin-shell factor
\be
\phi_{\rm screened} \approx  -\frac{g}{4\pi M_{\rm Pl}}\frac{3\Delta {\cal R}}{{\cal R}} \frac{Me^{-m_{\rm min-out} r}}{r} + \phi_{\rm min-out}\,.
\label{phiscreened}
\ee
Note that both here and in the unscreened case below, $g/M_{\rm Pl}\equiv {\rm d A}/{\rm d}\phi|_{\phi_{\rm min-out}}$.

The suppression factor $\Delta {\cal R}/{\cal R}\ll 1$ can alternatively be understood intuitively as follows: deep inside the source, the contribution to the exterior profile from
infinitesimal volume elements are Yukawa-suppressed due to the large effective chameleon mass in the core. Only the contributions from within the
thin shell propagate nearly unsuppressed to an exterior probe.

Clearly, the thin-shell screening breaks down for sufficiently small objects. Imagine shrinking the source keeping the density fixed.  Eventually
the cost in gradient energy becomes prohibitively large, and the scalar field can no longer reach $\phi_{\rm min-in}$ in the core of
the object. The thin-shell screening goes away, and the exterior profile takes on its usual, unsuppressed form
\be
\phi_{\rm unscreened} \approx  -\frac{g}{4\pi M_{\rm Pl}} \frac{Me^{-m_{\rm min-out} r}}{r} + \phi_{\rm min-out}\,.
\label{noscreen}
\ee
The criterion for thin-shell screening to be effective is for the right-hand side of~(\ref{thin}) to be $\ll 1$, in which case the
exterior profile is screened and given by~(\ref{thin}). If this ratio happens to be $\gsim 1$, on the other hand, then the exterior profile is
unscreened and given by~(\ref{noscreen}).

The interaction between two large, screened objects is reduced by their two thin-shell factors, {\it i.e.} the effective coupling is
\be
g_{\rm eff}^2=g^2 \left(\frac{3\Delta {\cal R}}{{\cal R}}\right)_1\left(\frac{3\Delta {\cal R}}{{\cal R}}\right)_2\,.
\label{geff}
\ee
On the other hand, the force between a large, screened object and a smaller, unscreened object is reduced by only one thin-shell factor
(assuming that the force is probed outside the large body, so that the chameleon value is not drastically altered at the small object position).

\section{Chameleons Through Dimensional Reduction}
\label{chamdimred}
\ \ \ \ \
We are interested in a finding an embedding of a chameleon-type theory \cite{Khoury:2003aq,Khoury:2003rn}
within string models, which amounts to deriving a theory such as (\ref{symmetronlagrangian}) from a string compactification.
In fact, we have a very general situation of this type when we do a Kaluza-Klein (KK) dimensional reduction on a compact space, from $D$ dimension down to 
$d$ dimensions. 

When we do a KK reduction of a $D$-dimensional theory on a $d$-dimensional Minkowski space $M_d$ times a compact space $K$, we start with the KK 
expansion of fluctuations in spherical harmonics, generically of the type 
\be
(\Phi-\Phi_{\rm bgr})_{(M,\Lambda)}(x,y)=\delta \Phi_{(M,\Lambda)}(x,y)=\sum_{n,I}\phi^{I(n)}_M(x)Y_\Lambda^{I(n)}(y),
\ee
where the spherical harmonics $Y_\Lambda^{I(n)}(y)$ are eigenfunctions of the kinetic operator on the compact space $\Delta_{K,y}$ (here the total 
kinetic operator of $\Phi_{(M,\Lambda)}(x,y)$ is $\Delta_{M\times K;x,y}=\Delta_{M;x}+\Delta_{K;y}$), $(M,\Lambda)$ are indices for the representation
of the Lorentz group on $M_d$ and $K$, respectively, and $I$ are indices for the representation of the symmetry group of $K$ at level $n$.

This KK expansion is linear ($\delta \Phi$ is linear in $\phi^{I(n)}$) and exact to all orders in perturbations, being nothing but a generalization
of the usual Fourier expansion on a circle, $\phi_M(x,y)=\sum_n \phi_M^{(n)}(x)e^{\frac{2\pi iny}{R}}$, with the ``spherical harmonics" $e^{\frac{2\pi i
ny}{R}}$ being eigenfunctions of $\d_y^2$ (where $\Delta_{x,y}=\Delta_x+\d_y^2$).

The KK {\em reduction} is then the truncation of the KK expansion to the ``zero-modes" $\{ \phi_M^{I(0)}(x)\}$. In general it can be inconsistent, {\it i.e.} 
we can have terms in the action linear in $\phi^{I(n)},n>0$, preventing us from putting them to zero, since it will not solve the $\phi^{I(n)}$ 
equations of motion. While consistent truncation is not a necessary requirement, it does simplify things, as we are allowed to work with a finite 
number of fields $\{\phi_M^{I(0)}(x)\}$ only. Sometimes we can write a nonlinear KK {\em reduction} ansatz instead, equivalent to a nonlinear 
redefinition of the linear fields $\{\phi_M^{I(n)}\}$, such that a consistent truncation to $\{\phi_M^{I(0)}(x)\}$ is possible and/or such 
that the $d$-dimensional action assumes a known form, like a supergravity action. For the metric, 
the nonlinear ansatz generically is of the type (see for instance \cite{Nastase:2000tu})
\bea
\nonumber
{\rm d}s_D^2&=&R^2{\rm d}s_d^2+g_{mn}{\rm d}x^m{\rm d}x^n+\ldots \\
\nonumber
&\equiv & g^{(D)}_{MN}{\rm d}x^M{\rm d}x^N\\
{\rm d}s_d^2&=&g_{\mu\nu}{\rm d}x^\mu {\rm d}x^\nu\,,
\label{nonlinearKKmetric}
\eea
where the (dimensionless) conformal factor is given by 
\be
R= \Delta^{-\frac{1}{d-2}}\,;\;\;\;\; \Delta=\sqrt{\det g_{mn}}\,.
\label{R}
\ee
Often one considers the linear KK expansion and the associated (possibly inconsistent) 
linear KK {\em reduction} ansatz, in which case one has a constant $R$ and would miss the effect described here.

Also, although there is no theorem saying that a full nonlinear KK {\em reduction} ansatz should exist, in all known cases of reducing a 
supergravity theory on a compact space, if there is a corresponding supergravity with the correct symmetries in the lower dimension, then 
there is a nonlinear KK ansatz which embeds the lower dimensional supergravity in Einstein frame into the higher dimensional action 
(see for instance the case of the $AdS_7\times S^4$ reduction in \cite{Nastase:1999cb,Nastase:1999kf}).

In these cases when there is a nonlinear ansatz for embedding, the matter action has a simple form in the Einstein frame, namely the matter 
part of the lower dimensional supergravity action. But firstly, as we mentioned there is no general theorem saying that this always happens, 
and secondly, there is 
no general way to decide whether the matter fields of the linear ansatz (coupling to the Jordan frame metric) or the nonlinear ansatz
(coupling to the Einstein frame metric) correspond to the matter we observe in our 4 dimensions. 

In any case, the nonlinear KK reduction ansatz for the metric, given in (\ref{nonlinearKKmetric}), is always valid, since it just 
reflects the rescaling needed to bring the $D$-dimensional Einstein term into the $d$-dimensional Einstein term:
\be
M_{D}^{D-2}\int {\rm d}^Dx\sqrt{G}R^{(D)}=M_{d}^{d-2}\int {\rm d}^dx\sqrt{g}R^{(d)}\,,
\ee
where $M_D$ and $M_d$ are the $D$- and $d$-dimensional Planck mass, respectively.
For the matter action, on the other hand, we have no {\it a priori} guidelines as to how the lower dimensional action should appear.
We will therefore simply choose a linear ansatz, that is, the $d$-dimensional matter fields will  couple to the $D$-dimensional Jordan frame metric, just like the $D$-dimensional matter fields in 
the $D$-dimensional action.
In the notation of~(\ref{conf}), this means  that ${\rm d}s^2_d=g_{\mu\nu}{\rm d}x^\mu {\rm d}x^\nu$, ${\rm d}s_D^2=\tilde g_{\mu\nu}{\rm d}x^\mu {\rm d}x^\nu+...$, and so
\be
A^2(\phi)=R^2= \Delta^{-\frac{2}{d-2}}= (\det g_{mn})^{-\frac{1}{d-2}}\,.
\ee
Focusing on the physical case of $d=4$ and the choice $D=10$ relevant to string theory, we therefore have the relations:
\bea
\nonumber
{\rm d}s^2_D&=&R^2{\rm d}s_4^2+g_{mn}{\rm d}x^m {\rm d}x^n=\Delta^{-1} {\rm d}s_4^2 +g_{mn}{\rm d}x^m {\rm d}x^n\,;\\
\nonumber
\Delta&=& \sqrt{\det g_{mn}} = V_6 M_{\rm Pl}^6 = r^6M_{\rm Pl}^6\,;\\
\nonumber
A^2(\phi)&=& R^2 = \Delta^{-1} \,;\\
R&=& \Delta ^{-\frac{1}{2}} = \frac{1}{\sqrt{V_6M_{\rm Pl}^6}} =  \frac{1}{r^3M_{\rm Pl}^3}\,,
\label{keyrelations}
\eea
where $V_6$ and $r$ denote respectively the volume and (average) radius of the 6 dimensional compact manifold.
Note that $R$ and $\Delta$ are both dimensionless, and $M_{\rm Pl} \equiv (8\pi G_{\rm N})^{-1}$ is the 4d Planck scale.

To compare with chameleon theories, we still need to go from the variable $R$ to the canonical scalar field $\phi$, with action as in~(\ref{symmetronlagrangian}). 
The exact form of the kinetic term for $R$, or $\Delta=\sqrt{\det g_{mn}}$, depends on the details of the compactification, but in general 
it will be such that the canonical scalar $\phi$ is of the form
\be
\phi= \frac{M_{\rm Pl}}{g}\ln \frac{R}{R_*} \,,
\label{phireln}
\ee
where, as we will confirm shortly, $g$ is a dimensionless coupling determining the strength of the scalar-mediated force, 
consistent with the notation of~(\ref{phiscreened}) and~(\ref{noscreen}). Meanwhile, $R_*$ is a reference field value that will be fixed shortly.
In the case of the $AdS_7\times S^4$ compactification, for instance,
we have the decomposition of $\Delta$ in terms of scalars as 
$\Delta^{-6/5}=Y^AT_{AB}Y^B$ and $T^{AB}={{\Pi^{-1}}_i}^A{{\Pi^{-1}}_i}^B$, with the kinetic term for these scalars given by
$-\frac{1}{2}P_{\a ij}P^{\a ij}$, where $P_{\a ij}$ is the symmetric part (at zero gauge field) of ${{\Pi^{-1}}_i}^A\d_\a{\Pi_A}^k\delta_{kj}$ (see \cite{Nastase:1999cb,Nastase:1999kf} for details and notation).

\begin{figure} %  figure placement: here, top, bottom, or page
   \centering
   \includegraphics[width=4.0in]{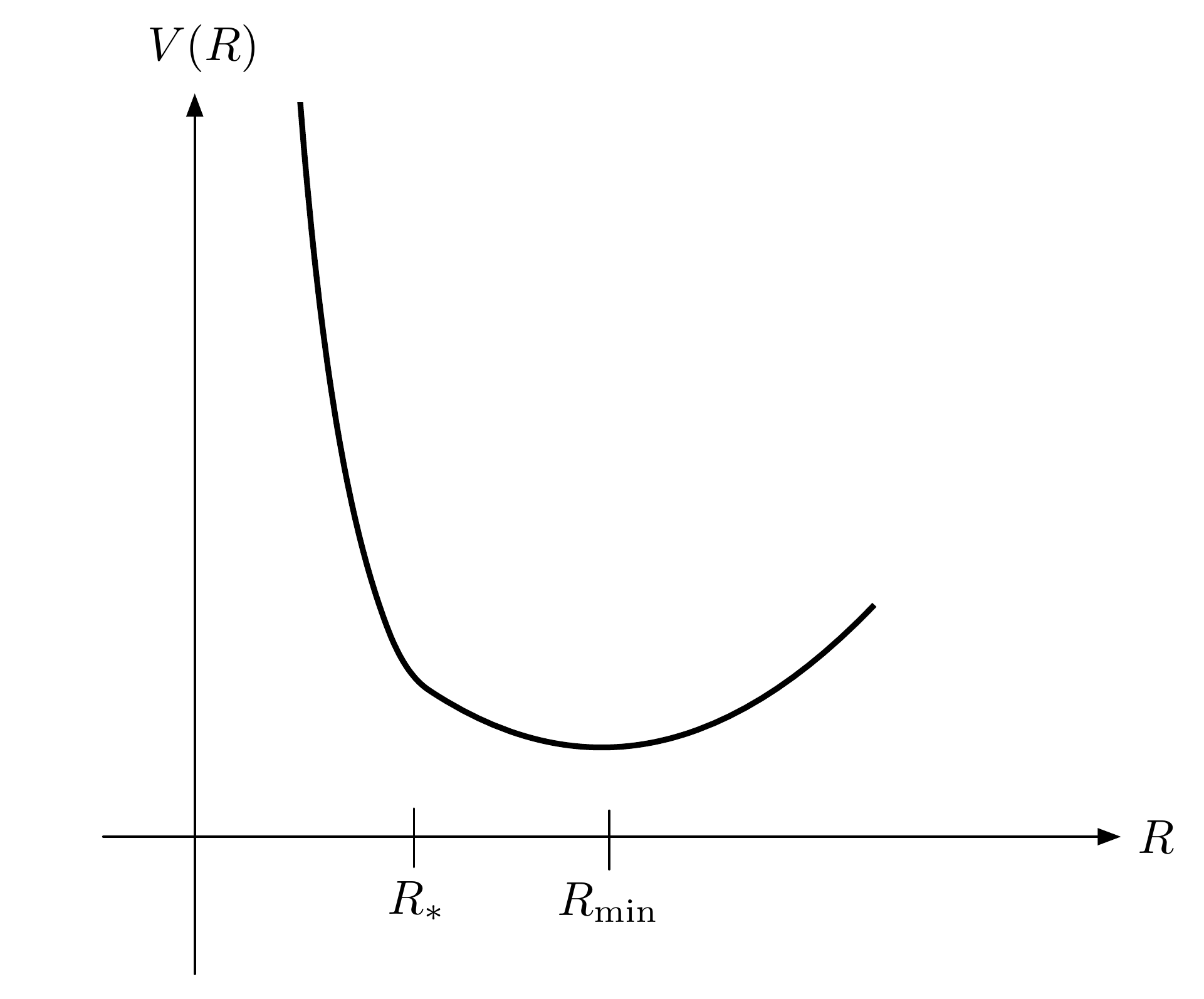}
   \caption{Our phenomenological potential consists of an exponentially steep part~(\ref{steep}) for $R < R_*$ and a quadratic form~(\ref{V(R)quad})
   for $R > R_*$, with minimum at $R = R_{\rm min}$.}
   \label{potentialsketch}
\end{figure}

In general there will be a potential for $R$, since the volume of the compact space will generically need stabilization. This gives
the $V(\phi)$ potential for the canonical scalar in (\ref{symmetronlagrangian}). We will try to be more general by first studying a very general toy potential expected to arise in such a situation.  We then move on to a specific stringy example in Sec.~\ref{egkklt}.  The toy potential consists of two regimes: $i)$ an approximately quadratic region around some minimum value $R_{\rm min}$; $ii)$ a steep exponential form starting at $R_\ast$, with $0<R_\ast<R_{\rm min}$. This is sketched in Fig.~\ref{potentialsketch}.

The exponential at $R_\ast>R>0$ is the simplest approximation to a steep wall at $R_\ast$, and reflects the fact that we cannot have $R=0$ since that would mean decompactification and the metric (\ref{nonlinearKKmetric}) would become singular, and 
we assume volume stabilization. The choice of a generic $R=R_*$ as the start of the steep exponential (as opposed to  $R_*$ close to zero) is a 
choice of model.
It is natural to fix the $R_*$ introduced in~(\ref{phireln}) as the $R_\ast$ in our toy potential, since it is the field value delineating the two regimes,
so that $R_\ast$ corresponds to $\phi = 0$. This general form is a good approximation to the KKLT potential, for instance, as discussed
in Sec.~\ref{egkklt}.  

The quadratic regime around $R_{\rm min}$ reflects the fact that the volume modulus is stabilized at $R_{\rm min}$.  For $R\gsim R_*$, we therefore approximate the potential as quadratic
\be
V(R)=M_{\rm Pl}^4\left[-\a (R-R_*)+\b (R-R_*)^2\right]\,,
\label{V(R)quad}
\ee
where $\a,\b>0$ are both dimensionless. Note that the only assumption here is the fact that $\a,\b>0$, which is necessary for the 
existence of a minimum. The minimum is located at
\be
R_{\rm min} = R_*+\frac{\a}{2\b}\,,
\label{Rmin}
\ee
corresponding to the volume of the extra dimensions stabilized in vacuum. For $R \lsim R_*$,
meanwhile, where the potential is approximated by a steep exponential rise,
a useful functional form (which approximates the KKLT potential discussed in Sec.~\ref{egkklt}) is
\be
V(R)= M_{\rm Pl}^4\, v\, \left[e^{\gamma (R^{-k}-R_*^{-k})}-1\right]\,,
\label{steep}
\ee
with $\gamma,k>0$ and $v>0$ all dimensionless. Of course, $\gamma<0, k<0$ would achieve the same result, but we stick to $\gamma,k>0$ since 
the KKLT potential has this form (with $k=4/3$).
The scale $R_*$ at which the behavior of the potential changes can be fixed in terms of the other parameters by matching ${\rm d}V/{\rm d}R$ 
for $R>R_*$ and $R<R_*$,
\be
\alpha = v\;\frac{k\gamma}{R_*^{k+1}}\,.
\label{matching}
\ee

The effective potential~(\ref{Veff}), including the coupling to matter density, is then given by\footnote{Note that
the density term is linear in $R$, hence this justifies using this field variable in defining the potential, as we did in~(\ref{V(R)quad}) and~(\ref{steep}).}
\be
V_{\rm eff}(\phi)=V(\phi) + \rho \frac{R}{R_*} = V(\phi)+\rho e^{g\phi/M_{\rm Pl}}\,.
\ee
Note that we have normalized the coefficient of the density term to be unity at $R=R_*$, corresponding to $A(\phi = 0) = 1$, without loss of generality. 
Any other choice could of course be absorbed in a rescaling of $\rho$. This particular choice singles out $\rho_*$ ---
the density corresponding to $R_*$ --- as the physical matter density, whereas other values of $\rho$ must be
multiplied by $R/R_*$ to obtain physical, Einstein-frame densities. In practice, we will see that $R$ varies little over
the relevant range of densities, hence there is almost no difference in taking $R_*$, or the minimum value of $R_*+\a/2\b$, or any other value of interest, as the normalization scale. 
 Note that $A(\phi) = e^{g\phi/M_{\rm Pl}}$ matches
the fiducial coupling function assumed in Sec.~\ref{generic}, confirming $g$ as the coupling parameter.

\section{Example: KKLT Potential}
\label{egkklt}
\ \ \ \
The potential introduced by KKLT~\cite{Kachru:2003aw} is in fact an example of the type discussed in the previous section. The 
relevant scalar field in this case is the complexified volume modulus $\varrho$, whose imaginary part gives the volume dependence through
\footnote{Note that in KKLT, the scale appearing in $\rho$ is not the 4d Planck scale, but rather the 10d string scale $M_s=(2\pi\sqrt{\a'})^{-1}$,
so $\sigma=M_s^4r^4/g_s=M^4r^4/(2\sqrt{\phi}g_s)$, where $M$ is the 10d Planck scale. Since we will see that the experimental constraints require 
large extra dimensions $r$, this makes a big difference. $M$ is of course fixed, and $M_{\rm Pl}$ is given by experiment, so the difference 
between the two definitions can be absorbed by rescaling the prefactor $a$ appearing in the nonperturbative superpotential below by a large number.
In the following we will continue to refer to such a model as KKLT, though obtaining $a$ of order 1 as we will assume from now on, 
as opposed to very small, is problematic for KKLT.}
\be
\sigma\equiv {\rm Im}\,\varrho = \frac{M_{\rm Pl}^4r^4}{g_s} = \frac{R^{-4/3}}{g_s}\,,
\label{sigmaKKLT}
\ee
where $g_s$ is the string coupling.
In the last equality we have assumed a normalization where $\Delta=\sqrt{\det g_{mn}}$ equals the volume of the extra dimensions in Planck units, $V_6M_{\rm Pl}^6$, as in Sec.~\ref{chamdimred} --- see the last of~(\ref{keyrelations}). The ${\cal N}=1$ supersymmetric set-up is obtained using a Klebanov-Strassler throat~\cite{Klebanov:2000hb} glued onto half of a Calabi-Yau space $CY_3$, which gives the tree-level K\"ahler potential for $\varrho$
\be
K=-3\ln [-i(\varrho-\bar\varrho)]=-3\ln [2\,{\rm Im}\varrho] \,,
\label{kahlerpot}
\ee
and a superpotential of the type
\be
W=W_0+Ae^{ia\varrho}\,,
\label{superpot}
\ee
with the constant and exponential parts in $W$ perturbative and non-perturbative, respectively.
In KKLT, the origin of the nonperturbative superpotential piece is two-fold: 1) Euclidean D3-brane instantons wrapping 4-cycles in the 
compact space~\cite{Witten:1996bn} (the 1-instanton action gives a factor of $e^{-V_D}$, where $V_D=2\pi{\rm Im}\tilde \varrho
\equiv 2\pi M_{s}^4r^4/g_s$ is the volume in Planck units of a 4-dimensional space wrapped by the instantons, 
thus giving $a=2\pi(M_s/M_{\rm Pl})^4$, whereas the prefactor $A$ is the one-loop determinant of fluctuations around the instanton), 
and 2) $N_c$ D7-branes wrapping 4-cycles and giving a ${\cal N}=1$ $SU(N_c)$ 
gauge theory. (Gluino condensation in this gauge theory gives a superpotential with $a=2\pi/N_c(M_s/M_{\rm Pl})^4$ and $A=N_cM_s^3$, with 
$M_s$ being the cut-off scale, {\it i.e.} the string scale, below which SQCD is valid.) 

From~(\ref{kahlerpot}) and~(\ref{superpot}), one obtains the supersymmetric potential\footnote{Note that $M_{\rm Pl}$ factorizes in~(\ref{VKKLT}) since $W$ has mass-dimension 3
while $\varrho$ is dimensionless.}
\be
V=\frac{e^{-a {\rm Im} \varrho}}{2M_{\rm Pl}^2}\left[\frac{A^2a^2}{3}( {\rm Im} \varrho)^{-1}e^{-a {\rm Im} \varrho}+Aa( {\rm Im} \varrho)^{-2}(Ae^{-a {\rm Im} \varrho}+W_0)\right]\,,
\label{VKKLT}
\ee
and the kinetic term for $\varrho$ 
\be
g_{i\bar j}D_\mu \phi^i (D^\mu \phi)^{\bar j}=\frac{3M_{\rm Pl}^2}{(2{\rm Im} \varrho)^2}\d_\mu \varrho (\d^\mu \varrho)^*\,.
\ee
Thus the canonically normalized field is $\phi=\pm M_{\rm Pl}\sqrt{3/2} \ln {\rm Im} \varrho=\mp 4M_{\rm Pl}/\sqrt{6}\ln R+\phi_0$.
Defining Im$\varrho=cR^{-4/3}$, the supersymmetric KKLT potential is
\be
V=\frac{e^{-acR^{-4/3}}}{2M_{\rm Pl}^2}\left[\frac{A^2a^2}{3c}R^{4/3}e^{-acR^{-4/3}}+\frac{Aa}{c^2}R^{8/3}\Big(W_0+Ae^{-acR^{-4/3}}\Big)\right]\,.
\label{poten}
\ee
Though $c = 1/g_s$ in reality, as seen from~(\ref{sigmaKKLT}), by trivial redefinitions of parameters we will henceforth assume $c = 1$ for simplicity.

Finally, a nonsupersymmetric (de Sitter) minimum is generated by adding an anti-D3-brane at the tip of the KS throat, giving a 
potential term $D/({\rm Im}\varrho)^3=D'R^4$. As we will see, however, over the field range of interest this term is well-approximated by
a constant, since its variation is negligible compared to that of the exponentials. We can effectively shift the entire
potential to obtain a desired cosmological constant. Therefore we will not worry about the cosmological constant from now on, and
we will restrict our attention to the supersymmetric potential~(\ref{poten}).

\begin{figure} %  figure placement: here, top, bottom, or page
   \centering
   \includegraphics[width=4.0in]{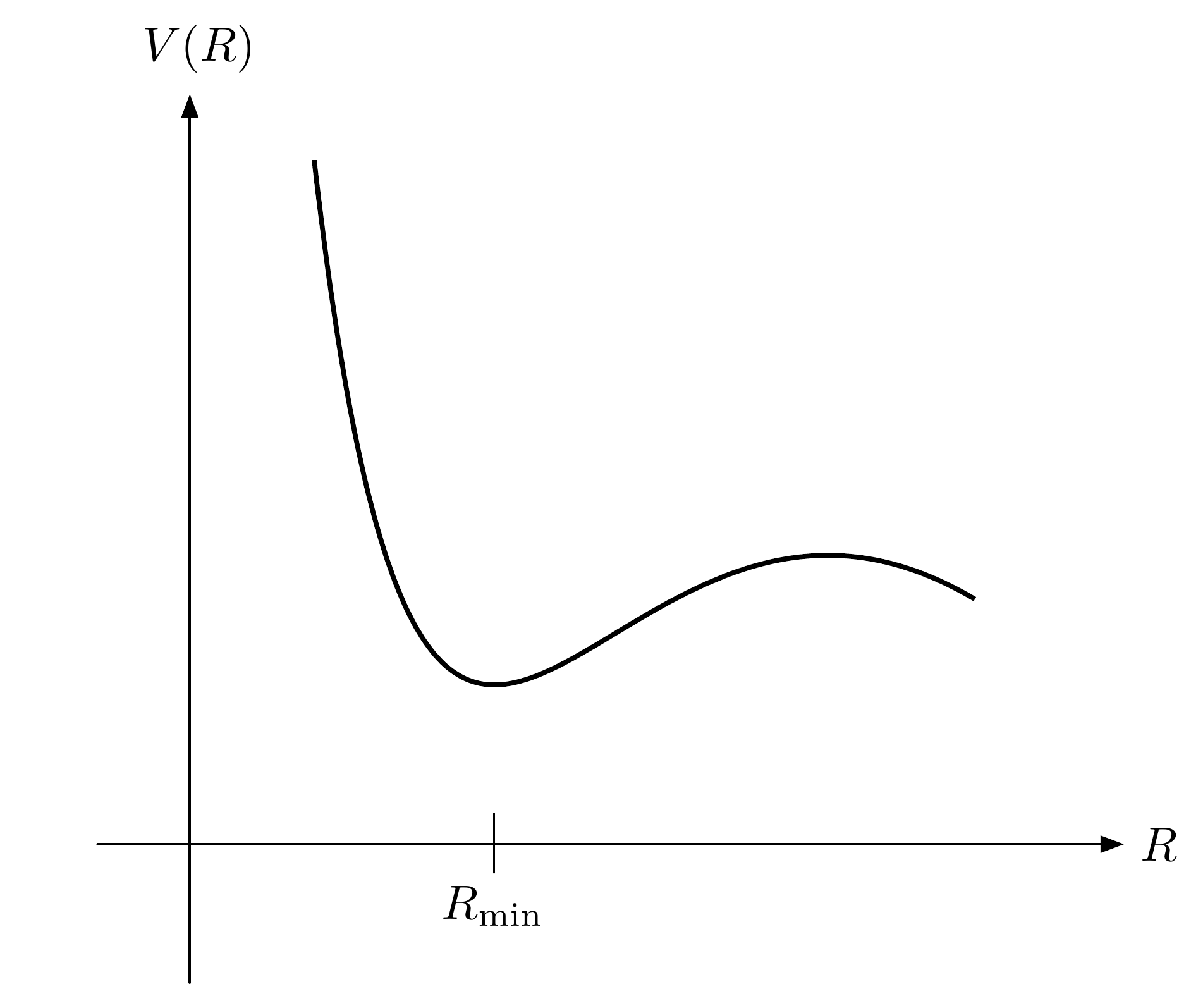}
   \caption{Sketch of the KKLT potential~(\ref{poten}) with $a<0$. The region of interest for the chameleon mechanism is $R \leq R_{\rm min}$.}
    \label{KKLTfig}
\end{figure}

The KKLT potential~(\ref{poten}) satisfies many of the key features of the generic potential discussed in Sec.~\ref{chamdimred}.
For KKLT, one needs that the 1-instanton action be large, {\it i.e.} $|a\varrho|\gg 1$, while the 1-loop determinant 
prefactor $A$ in Case 1) can in principle be suppressed. Both of these conditions are required in our case, as we
will see shortly. An important difference, however, is that we need $a<0$, unlike the explicit KKLT case. Hence
our desired potential is not strictly realized in the original KKLT model. Nevertheless, superpotentials containing the needed positive exponentials, 
$e^{-i|a|\varrho}=e^{|a|{\rm Im}\varrho+...}$, are easily found in string theory --- even within the KKLT context it was considered in 
\cite{Abe:2005rx}, including a highly suppressed prefactor $A$.\footnote{The superpotential with $A=Ce^{-m_9c<S>}$ is obtained by including gluino 
condensation on an extra D9-brane with magnetic flux in the KKLT scenario, where $2\pi S=e^{-\phi}-ic_0$ is the dilaton modulus. Here $c=8\pi^2/N_9$
and the coupling function on the D9-brane is $1/g_{D9}^2=|m_9 {\rm Re}S-w_9{\rm Re T}|$, with $T=-i\varrho$. Since as we will see later we need 
${\rm Im }\varrho\sim \log A\gg 1$, it is possible to arrange for a coupling $g_{D9}$ of order 1 if one desires.
However, again with our definition of $\sigma$, $a$ is of order $(M_s/M_{\rm Pl})^4$, as opposed to order 1 as we assumed here.}
See~\cite{Quevedo:1996sv,Burgess:1997pj} for other examples.\footnote{In particular, in the review \cite{Quevedo:1996sv} it is explained that
by imposing T-duality invariance (and modular invariance) on gaugino condensation superpotentials obtained for tori compactifications  
with dilaton modulus $S$ 
and volume modulus $T$, one obtains generically superpotentials of the form $W(S,T)\sim \eta(iT)^{-6}\exp (-3S/8\pi b)$, with $\eta(x)$ the 
Dedekind eta function. At large volume Re $T\rightarrow \infty$, corresponding to the regime we will need later, we obtain $W\propto e^{\pi T/2}$,
with a coefficient which is again exponentially small in the dilaton modulus, $\exp(-3S/8\pi b)$, with $b$ a renormalization group factor of order
1. Thus the needed superpotential term is now obtained in the limit we will be using. The same comment applies about the factor $a$ in the 
exponent as before.}
In the following, we will continue to refer to this model as KKLT, even though the explicit nonperturbative mechanisms considered
in~\cite{Kachru:2003aw} do not yield $a<0$, nor $a\sim {\cal O}(1)$. The potential~(\ref{poten}) with $a<0$ is sketched in Fig.~\ref{KKLTfig}. 

We now derive a few properties of the KKLT potential that will be useful later on. The minimum of the potential in terms of $\sigma={\rm Im} \varrho$ satisfies $DW=0$, which gives\footnote{In the original KKLT paper~\cite{Kachru:2003aw}, note that $\sigma_{\rm min}$ is instead denoted by $\sigma_c$.}
\be
W_0=-Ae^{-a\sigma_{\rm min}}\left(1+\frac{2}{3}a\sigma_{\rm min}\right) \approx \frac{2}{3} A|a| \sigma_{\rm min} e^{|a|\sigma_{\rm min}}\,,
\label{minimumcond}
\ee
where in the last step we have used the fact that $a < 0$ and assumed $|a|\sigma_{\rm min} \gg 1$. In this limit, we can solve
approximately for the field value at the minimum:
\be
\sigma_{\rm min} \approx \frac{1}{|a|} \ln\frac{3W_0}{2A}\,.
\label{sigminW0}
\ee
The value of the potential at the minimum is then given by
\be
V_0=-\frac{A^2a^2e^{2|a|\sigma_{\rm min}}}{6M_{\rm Pl}^2\sigma_{\rm min}} \approx - \frac{3W_0^2}{8M_{\rm Pl}^2\sigma_{\rm min}^3}\,.
\label{minimumpot}
\ee
Meanwhile, the curvature of the potential around the minimum is, again for $|a|\sigma_{\rm min} \gg 1$,
\bea
\left.\frac{{\rm d}^2V}{{\rm d}\sigma^2}\right\vert_{\rm min} &=&\frac{V_0}{\sigma_{\rm min}^2}\left[6+4a\sigma_{\rm min}+a^2\sigma^2_{\rm min} \right]+\frac{\left[Aa\sigma_{\rm min}e^{-a\sigma_{\rm min}}\right]^2}{6\sigma_{\rm min}^5}
\left(8+9a\sigma_{\rm min}+3a^2\sigma_{\rm min}^2\right)\cr
\nonumber
&\simeq &\frac{3}{4}\frac{W_0^2}{M_{\rm Pl}^2\sigma_{\rm min}^5}a^2\sigma_{\rm min}^2 \\
&=& 2a^2(-V_0)\,.
\label{sigmamass}
\eea

In order to check whether (\ref{poten}) works as an example of the type from the previous section, let us consider the KKLT potential around its minimum. Combining (\ref{poten}), (\ref{minimumcond}), (\ref{minimumpot}), we obtain the exact result
\be
V_{\rm KKLT}(\sigma)-V_0=\frac{A^2a}{2M_{\rm Pl}^2}\left[\frac{a}{3}\Big(\frac{e^{-2a\sigma}}{\sigma}+\frac{e^{-2a\sigma_{\rm min}}}{\sigma_{\rm min}}-\frac{2\sigma_{\rm min}}
{\sigma^2}e^{-a(\sigma+\sigma_{\rm min})}\Big)+\frac{e^{-a\sigma}}{\sigma^2}(e^{-a\sigma}-e^{-a\sigma_{\rm min}})\right]\,.
\ee
Since we need $|a|\sigma\sim |a| \sigma_{\rm min} \gg 1$ in order to have a steep exponential, we can, in the region of interest, neglect the last term of the potential
\be
V_{\rm KKLT}(\sigma)-V_0\simeq\frac{A^2a^2}{6M_{\rm Pl}^2}\left[\frac{e^{2|a|\sigma}}{\sigma}+\frac{e^{2|a|\sigma_{\rm min}}}{\sigma_{\rm min}}-\frac{2\sigma_{\rm min}}{\sigma^2}e^{
|a|(\sigma+\sigma_{\rm min})}\right]\,.
\label{approxi}
\ee
It is clear that we can obtain a steep exponential rise, as desired, for $\sigma > \sigma_*$ where $|a|(\sigma_*-\sigma_{\rm min})\sim 1$. In this
regime, the potential can be approximated by the leading exponential 
\be
V_{\rm KKLT}(\sigma)-V_0\simeq \frac{A^2a^2}{6M_{\rm Pl}^2\sigma_*}\left[e^{2|a|\sigma}-e^{2|a|\sigma_*}\right]\,.
\label{approx2}
\ee
It is also easy to see that, since $|a|(\sigma_*-\sigma_{\rm min} )\sim 1$ and the minimum is made by combining exponentials, the quadratic approximation in 
terms of $R$ is not a good one for the KKLT potential. Nevertheless, as we will discuss later, all we need is the existence of a minimum of the potential 
for $R>R_*$, so in fact KKLT works as an example of the general set-up described in Sec.~\ref{chamdimred}. Therefore, when fixing the parameters we will 
be guided by the KKLT example. For instance, we will apply mostly the value $k=4/3$ in~(\ref{steep}), as implied by~(\ref{approx2}) and the relation~(\ref{sigmaKKLT}) between $\sigma$ and $R$.

\section{Constraints on the model and phenomenological implications}
\label{consastro}
\ \ \ \ 
In section \ref{generic} we have seen how chameleon theories operate in the generic case. Now we apply 
the discussion to the phenomenological potential of Sec.~\ref{chamdimred}, described by~(\ref{V(R)quad}) for $R>R_*$ and~(\ref{steep}) for $R<R_*$. We will first obtain the tightest constraints on the parameters
of  the model, namely the effective exponent $\gamma k R_*^{-k}$ and the parameters $\a$ and $\b$, arising from laboratory searches for a fifth force.
We will argue that all other constraints, such as from solar system tests, automatically follow from laboratory bounds in this case. At the end of the section, we discuss 
how the constraints thus derived are affected if the potential instead consists of the {\it sum} of~(\ref{V(R)quad}) and~(\ref{steep}).

\subsection{Laboratory Constraints}
\label{labcons}
\ \ \ \
Because the chameleon couples to matter with gravitational strength, the tightest constraint comes from tests of gravity in the laboratory, specifically fifth
force searches. As detailed in~\cite{Khoury:2003aq}, these experiments are performed in vacuum, where the chameleon acquires a Compton wavelength set by
the size of the vacuum chamber: $m^{-1}_{\rm vac} \simeq {\cal R}_{\rm vac} \sim 10\;{\rm cm}-1\;{\rm m}$. On these scales, Hoskins {\it et al.}~\cite{Hoskins:1985tn} constrains
the fifth force to be weaker than gravity by a factor of $10^{-3}$, that is,
\be
g_{\rm eff}^2 \;\lsim\; 10^{-3}\,.
\ee
Thus the test masses used in these experiments must be screened, with a thin shell satisfying
\be
\left(\frac{3\Delta {\cal R}}{{\cal R}}\right)_{\rm test\;mass} = \frac{\phi_{{\rm vac}}-\phi_{\rm test \; mass}}{2gM_{\rm Pl}\Phi_{\rm test \; mass}} \;\lsim\; 10^{-3/2}\,,
\label{testmass}
\ee
where $\phi_{\rm vac}$ is the field value around which the chameleon mass is $m^{-1}_{\rm vac}\simeq 10\;{\rm cm}-1\;{\rm m}$,
and $\phi_{\rm test \; mass}$ is the field value corresponding to $\rho_{\rm test\;mass}\sim 10\;{\rm g}/{\rm cm}^3$.
The test masses used in~\cite{Hoskins:1985tn} are estimated to have $\Phi_{\rm test \; mass}\simeq 10^{-27}$~\cite{Khoury:2003aq}.
Assuming $g\sim {\cal O}(1)$, we obtain
\be
\phi_{{\rm vac}}-\phi_{\rm test \; mass} \;\lsim\; 10^{-29}\; M_{\rm Pl}\,.
\label{vaccons}
\ee
Because of the small Newtonian potential involved, this is by far the most stringent constraint on our model.

We can straightforwardly translate this to the parameters of our potential by assuming that the field values lie on the steep exponential part~(\ref{steep}),
corresponding to $R < R_*$. Because the potential is steep, all values of interest satisfy $R_*-R \ll R_*$, hence~(\ref{phireln}) implies
\be
\phi \simeq \frac{M_{\rm Pl}}{g} \frac{R-R_*}{R_*}\,.
\label{phirelnapprox}
\ee
The mass around any field value on this steep part is dominated by differentiation of the exponent,
\be
m^2 \simeq \left(\frac{{\rm d}R}{{\rm d}\phi}\right)^2 \frac{{\rm d}^2V}{{\rm d}R^2} \simeq g^2M_{\rm Pl}^2 v\,\left(\frac{\gamma k}{R^{k}}\right)^2e^{\gamma(R^{-k}-R_*^{-k})} \,.
\label{mapprox}
\ee
In particular, for the vacuum case we obtain
\be
\frac{R_{\rm vac} - R_*}{R_*}\simeq \frac{R_*^k}{k\gamma}\log\left(\frac{g^2M_{\rm Pl}^2v}{m^2_{\rm vac}}\left(\frac{k\gamma}{R_*^k}\right)^2\right)\,.
\label{Rvac}
\ee
Meanwhile, in this regime the effective potential is minimized by
\be
\rho\simeq \frac{\gamma k}{R_*^k}M_{\rm Pl}^4 v\,e^{\gamma (R^{-k}-R_*^{-k})}\,,
\label{rhoRapprox}
\ee
which, for the test masses of interest, implies
\be
\frac{R_{\rm test \; mass} - R_*}{R_*}\simeq \frac{R_*^k}{k\gamma}\log\left(\frac{M_{\rm Pl}^4v}{\rho_{\rm test\; mass}}\frac{k\gamma}{R_*^k}\right)\,.
\label{Rtest}
\ee
Subtracting~(\ref{Rtest}) from~(\ref{Rvac}), we obtain a relation for $R_{\rm vac} - R_{\rm test\; mass}$
\be
\frac{R_{\rm vac} - R_{\rm test \; mass}}{R_*}\simeq \frac{R_*^k}{k\gamma} \log\left(\frac{g^2\rho_{\rm test\; mass}}{m^2_{\rm vac} M_{\rm Pl}^2} \frac{k\gamma}{R_*^k}\right)\,.
\ee
For the relevant values $\rho_{\rm test\;mass}\sim 10\;{\rm g}/{\rm cm}^3 \simeq 10^{-90}\;M_{\rm Pl}^4$ and $m_{\rm vac}\sim (10\;{\rm cm})^{-1}
\simeq 10^{-33}\;M_{\rm Pl}$,
and assuming $g\sim {\cal O}(1)$ as before, the constraint~(\ref{vaccons}) reduces to
\be 
\frac{R_*^k}{k\gamma} \left[ \log\left( \frac{k\gamma}{R_*^k}\right) - 24\right]\;\lsim\; 10^{-29}\,,
\ee
that is,
\be
R_*^k\;\lsim\; 10^{-30}\gamma k\,.
\label{Rkcond}
\ee
Since we will assume $\gamma, k\sim {\cal O}(1)$, this amounts to a constraint on $R_*$.

Finally, we should note that one might think we could avoid the lab constraints in a similar way to the symmetron case. There, the Compton wavelength
of the chameleon in the atmosphere is much larger than the size of a typical experiment, but also the effective coupling $g$ at the position of the 
experiments is driven to very small values due to the presence of the Earth nearby. In our case, we could realize the first condition, but not the 
second, since $g=M_{\rm Pl}{\rm d}A/{\rm d}\phi(\phi_{\rm atm})$ is ${\cal O}(1)$, unlike for the symmetron.

\subsection{Chameleon range in various environments}
\ \ \ \
We now derive bounds on the chameleon range locally, by assuming that the ambient matter density is sufficiently high for the field to lie
within $R < R_*$ throughout the solar system, where the potential is dominated by the steep exponential part~(\ref{steep}). In this case,
we can combine~(\ref{mapprox}) and~(\ref{rhoRapprox}) as follows
\be
m^2 \simeq g^2 \frac{\gamma k}{R_*^k}\frac{\rho}{M_{\rm Pl}^2} \;\gsim\; g^210^{30}\frac{\rho}{M_{\rm Pl}^2}\,,
\label{chamrange}
\ee
where the last step follows from~(\ref{Rkcond}). Assuming $g\sim {\cal O}(1)$, we thus find
\be
m \;\gsim\; 10^{15} \frac{\sqrt{\rho}}{M_{\rm Pl}} \,.
\ee
Recalling that $M_{\rm Pl} = 4.34\cdot 10^{-6}\;{\rm g}$, the range is
\be
m^{-1} \;\lsim\; \frac{0.2 \;{\rm mm}}{\sqrt{\rho[{\rm g}/{\rm cm}^3]}}\,.
\label{rangefinal}
\ee
For the Earth's mean density, $\rho_\oplus = 5.5\;{\rm g}/{\rm cm}^3$, this gives $m_\oplus^{-1} \;\lsim\; 0.1\;{\rm mm}$. For atmospheric density,
$\rho_{\rm atm} = 10^{-3}\;{\rm g}/{\rm cm}^3$, we find $m_{\rm atm}^{-1} \;\lsim\; 1\;{\rm cm}$. For the typical density within the solar system
($\rho_{{\rm solar}\; {\rm system}} \sim 10^{-24}\;{\rm g}/{\rm cm}^3$), $m^{-1}_{{\rm solar}\; {\rm system}}  \;\lsim\; 10^{5}\;{\rm km}$.
Thus the chameleon is relatively short-range for all relevant densities in the local environment. Nevertheless, it is worth emphasizing the
broad range of mass scales assumed, from $0.1\;{\rm mm}$ inside the Earth to $10^{5}\;{\rm km}$ in space. 

\subsection{Solar system tests}
\ \ \ \
Since the chameleon range is $\;\lsim\; 10^{5}\;{\rm km}$ in the solar system, constraints from planetary orbits around the
Sun are trivially satisfied. This distance scale is actually of the same order as the Earth-Moon distance, hence the relevant tests
come from Lunar Laser Ranging. However, a moment's thought reveals that the predicted departures from General Relativity
are minute, because both the Earth and the Moon are screened. In fact, it is straightforward to show that~(\ref{thin}) breaks down
in this case, as per the discussion below that equation, and the thin shell factors are instead given by
\bea
\nonumber
\left(\frac{3\Delta {\cal R}}{{\cal R}}\right)_{\oplus} &\simeq &\frac{1}{m_\oplus {\cal R}_\oplus} \;\lsim\; 10^{-11}\;;  \\
\left(\frac{3\Delta {\cal R}}{{\cal R}}\right)_{\rm Moon} &\simeq& \frac{1}{m_{\rm Moon} {\cal R}_{\rm Moon}} \;\lsim\; 10^{-10}\,.
\label{earththin}
\eea
Similarly for the Sun and other planets. The thin shell suppression also ensures that constraints from binary pulsars are satisfied.

\subsection{The galaxy and beyond}
\ \ \ \
The above analysis hinges on the Milky Way galaxy being screened, for otherwise the field value
in the solar system, $\phi_{{\rm solar}\; {\rm system}}$, would not be fixed by the local density. 
The thin shell condition for the galaxy is
\be
\left(\frac{3\Delta {\cal R}}{{\cal R}}\right)_{\rm G}  = \frac{\phi_{\rm cosmo} - \phi_{{\rm solar}\; {\rm system}}}{2gM_{\rm Pl}\Phi_{\rm G}} < 1\,,
\label{galaxyscreened}
\ee
where $\Phi_{\rm G}\sim 10^{-6}$. The case of interest is that the field value at cosmic density,
$\phi_{\rm cosmo}$, corresponds to $R_{\rm min} = R_* + \alpha/2\beta$, the minimum of the bare potential $V(R)$. 
Assuming that the displacement in $R$ is small, $|R_{\rm min} - R_{\rm solar system}| \ll R_*$, as we will confirm shortly,
such that~(\ref{phirelnapprox}) applies, we obtain
\be
\frac{R_{\rm min} - R_{\rm solar\; system}}{R_*} \;\lsim\; g^210^{-6}\,.
\label{RminRsolar}
\ee
With $R_{\rm solar\; system}\approx R_*$ and $g\sim {\cal O}(1)$, this implies
\be
\frac{\alpha}{\beta} \;\lsim\; 10^{-6}R_*\,,
\label{abcond2}
\ee
which is the desired condition on $\alpha/\beta$. In particular, the field displacement is indeed small\footnote{One could also consider the
case where the field already reaches $R_{\rm min}$ in the solar system, which would lead to a tighter constraint on $\alpha/\beta$ than~(\ref{abcond2}).}.

We can also derive a constraint on $\alpha$ alone as follows. The assumption made above of having $\phi_{\rm cosmo}$
correspond to $R_{\rm min}$ is only consistent provided that the cosmological density is low enough that the field does
not instead lie on the steep exponential part. In other words, the transition density $\rho_*$ for which $V_{\rm eff}$ is minimized at $R=R_*$
must be larger than the cosmological density
\be
\rho_*=-R_*\left.\frac{{\rm d}V}{{\rm d}R}\right\vert_{R_*} = \a R_* M_{\rm Pl}^4 \geq H_0^2M_{\rm Pl}^2 \,,
\label{rho*}
\ee
and hence
\be
\alpha \geq \frac{H_0^2}{M_{\rm Pl}^2R_*} \simeq 10^{-120}R_*^{-1}\,,
\label{alphaconstraint}
\ee
as desired. Combining this with~(\ref{abcond2}) also yields a constraint on $\beta$,
\be
\sqrt{\b}\geq 10^3 \frac{H_0}{M_{\rm Pl}R_*}\simeq  10^{-57} R_*^{-1} \,.
\label{betaconstraint}
\ee
In physical terms, this translates to a constraint on the mass of the chameleon for cosmological density (and lower, including in vacuum).
From~(\ref{V(R)quad}) and~(\ref{phirelnapprox}), we have
\be
m_{\rm cosmo} = \sqrt{2}gM_{\rm Pl} \sqrt{\beta}R_* \gsim 10^3H_0\,,
\label{mcosmo}
\ee
which means that the maximal range of the chameleon force is $\sim {\rm Mpc}$. This scalar field can therefore have important implications for structure formation~\cite{Schmidt:2008tn,Khoury:2009tk,Hui:2009kc,Wyman:2010jp,Chang:2010xh}.

To summarize, the key results are~(\ref{Rkcond}),~(\ref{alphaconstraint}),~(\ref{betaconstraint}), which constrain $R^k_*$, $\alpha$ and $\beta$, respectively. Given
$R^k_*$ and $\alpha$ satisfying these, the overall scale $v$ of the steep exponential potential is then fixed by~(\ref{matching}).

\subsection{Another perspective on the potential}
\ \ \ \
We can take two points of view on the potential for $R>R_*$ and $R<R_*$. The point of view taken until now was that both are approximations to 
the real potential, valid in each of the two regions. Alternatively, we can instead consider the potential as the sum of the steep exponential and the
quadratic form:
\be
V(R) = M_{\rm Pl}^4\left\{ v\, \left[e^{\gamma (R^{-k}-R_*^{-k})} -1\right]-\a (R-R_*)+\b (R-R_*)^2\right\}\,.
\label{Vsum}
\ee
 If we assume that~(\ref{matching}) still applies, then clearly the exponential will dominate for $R<R_*$, while the
quadratic piece will dominate for $R>R_*$. But we now argue for completeness that~(\ref{matching}) is not necessary --- all
approximations can still be satisfied.

In this case, the ${\rm d}^2V/{\rm d}R^2$ term in~(\ref{mapprox}) and~(\ref{chamrange}) would receive an additional $2\b M_{\rm Pl}^4$ contribution. In order for the above analysis to still be valid, this correction must be negligible for the relevant densities, 
\be
\b<\frac{\rho}{M_{\rm Pl}^4} \frac{\gamma k}{R_*^{k+2}}\,.
\ee
This constraint is tightest for $\rho_{{\rm solar}\; {\rm system}} \sim 10^{-24}\;{\rm g}/{\rm cm}^3\simeq 10^{-115}\;M_{\rm Pl}^4$, the lowest density assumed for
the steep exponential part. Combined with the upper bound~(\ref{betaconstraint}), we obtain
\be
\frac{10^{-114}}{R_*^2} < \b <\frac{10^{-115}}{R_*^2} \frac{\gamma k}{R_*^k} \,.
\label{betarange}
\ee
Since $\gamma k/R_*^k \;\gsim\; 10^{30}$ from~(\ref{Rkcond}), this allows for a broad range of values for $\beta$.

Similarly, ${\rm d}V/{\rm d}R$ in~(\ref{rhoRapprox}) would get an additional $M_{\rm Pl}^4\a $ contribution, which for consistency should be negligible for the relevant
densities: $\a<R_*^{-1} \rho/M_{\rm Pl}^4$. Focusing as before on $\rho_{{\rm solar}\; {\rm system}}$, and combining with~(\ref{alphaconstraint}),
we find the allowed range for $\alpha$:
\be
10^{-120}R_*^{-1} \;\lsim\; \alpha < 10^{-115} R_*^{-1} \,.
\label{alpharange}
\ee
Therefore, if $\alpha$ and $\beta$ fall within the above ranges, then all the constraints of Sec.~\ref{consastro} carry over to the total potential~(\ref{Vsum}).

\section{Applying the Analysis for the KKLT potential}
\label{KKLTapply}

\subsection{Applying the constraints}
\ \ \ \
In this Section we will translate the constraints derived for the phenomenological potential~(\ref{Vsum}) to the realistic example of the KLLT potential~(\ref{approxi}):
\be
V_{\rm KKLT}(\sigma)-V_0\simeq\frac{A^2a^2}{6m_{Pl}^2}\Big[\frac{e^{2|a|\sigma}}{\sigma}+\frac{e^{2|a|\sigma_{\rm min}}}{\sigma_{\rm min}}-\frac{2\sigma_{\rm min}}{\sigma^2}e^{
|a|(\sigma+\sigma_{\rm min})}\Big]\,.
\ee
The parameter $|a|$ is generically of order 1, both in the explicit nonperturbative mechanisms considered by KKLT and in the examples with $a<0$.
Since this parameter corresponds to $\gamma$ in our phenomenological potential, we also assume $\gamma \sim 1$. Meanwhile, from~(\ref{sigmaKKLT}) we note
that KKLT corresponds to $k=4/3$, in which case~(\ref{Rkcond}) reduces to $R_*\;\lsim\; 10^{-23}$. 

For concreteness, we will assume the limiting values $R_*\simeq 10^{-23}$, $\alpha/\beta \sim 10^{-6}R_*$ and $\beta \sim 10^{-114}/R_*^2$ allowed by~(\ref{Rkcond}),~(\ref{abcond2}) and~(\ref{betaconstraint}), respectively. From the last of~(\ref{keyrelations}), this immediately implies that the typical size of the extra dimensions is $r \simeq 10^{7.5} M_{\rm Pl}^{-1}$, corresponding to the Kaluza-Klein (KK) scale, is
\be
E_{\rm KK}\equiv r^{-1} \simeq 10^{11}\;{\rm GeV}\,.
\label{EKK}
\ee
Thus, with our choice of parameters, the KKLT potential naturally gives an intermediate KK scale for these extra dimensions.
\footnote{Note that this KK scale implies the 10d Planck scale is $M\simeq 10^{13}GeV$, and $M_s^4/M_{\rm Pl}^4\sim 10^{-23}$, so we would 
need an $a$ which is $10^{23}$ times larger than natural.}

Next we constrain $V_0$, the value of the potential at $\sigma_{\rm min}$, by noting that the curvature ${\rm d}^2V/{\rm d}\sigma^2_{\rm min}$ at the minimum
can be translated to the phenomenological potential through
\be
\left.\frac{{\rm d}^2V}{{\rm d}\sigma^2}\right\vert_{\rm min} = \left(\frac{{\rm d}R}{{\rm d}\sigma}(\sigma_{\rm min})\right)^2\frac{{\rm d}^2V}{{\rm d}R^2}(R_{\rm min})=\frac{9}{8} \b R_{\rm min}^2 R_{\rm min}^{8/3}M_{\rm Pl}^4\,.
\ee
Neglecting the small difference between $R_*$ and $R_{\min}$, and using the limiting value $R_*^{4/3} \simeq 10^{-30}$, 
we find ${\rm d}^2V/{\rm d}\sigma^2_{\rm min} \simeq 10^{-174}\;M_{\rm Pl}^4$. But comparing to~(\ref{sigmamass}) with $a\sim 1$ fixes $V_0$:
\be
|V_0| \simeq 10^{-174}\;M_{\rm Pl}^4\,.
\label{V0answer}
\ee
Note that this is not necessarily to be compared with the value of the cosmological constant, since in the end we must add to the
KKLT potential the SUSY-breaking term $D/\sigma^3=DR^4$, which over the field range of interest looks like a constant shift
of the potential to positive values. Substituting~(\ref{V0answer}) and $\sigma_{\rm min} \simeq R_*^{-4/3} \sim 10^{30}$ in the
relations~(\ref{sigminW0}) and~(\ref{minimumpot}), we can solve for $W_0$ and $A$:
\be
W_0 \simeq 10^{-42} M_{\rm Pl}^3\,; \qquad A \simeq W_0e^{-10^{30}} \sim M_{\rm Pl}^3e^{-10^{30}}\,.
\ee
Despite the very large number in the exponent for $A$, the superpotential term can be rewritten in a more natural form,
\be
Ae^{-a\sigma}=M_{\rm Pl}^3e^{|a|(\sigma-\sigma_0)}\,,
\label{sigma0}
\ee
with $\sigma_0\sim \sigma_{\rm min}\sim 10^{30}$.

\subsection{KKLT corrections to the constraints}
\ \ \ \
In the previous subsection we blindly applied the constraints on our phenomenological potential to $V_{\rm KKLT}$. However, as we already noted, 
the quadratic approximation for $R>R_*$ is not good for the latter. Nonetheless, all we need for our mechanism to work is a minimum for the potential, not
necessarily a simple quadratic form of the potential for $R>R_*$. In the following we will investigate the effect of using the full KKLT 
potential on the constraints we obtained before.

First note that the bound~(\ref{Rkcond}) on $R_*^k/\gamma k$ relied exclusively on the steep part of the potential, and thus 
remains valid. Similarly,~(\ref{chamrange})$-$(\ref{earththin}) still apply in this case. On the other hand, the constraint~(\ref{abcond2}) on $\alpha/\beta$,
which arose from demanding that the galaxy is screened, $3\Delta {\cal R}_{\rm G}/{\cal R}_{\rm G} <1$, did rely on the quadratic form near the minimum.
The more general condition, without specializing to the quadratic form, is given by~(\ref{RminRsolar})
\be
\frac{R_{\rm min} - R_{\rm solar\; system}}{R_*} \;\lsim\; 10^{-6}\,,
\label{rconstra}
\ee
where we have assumed $g\sim {\cal O}(1)$. We now translate this bound into a constraint on KKLT parameters. The left hand side can be written as
\be
\frac{3}{4}\frac{-a(\sigma_*-\sigma_{\rm min})}{-a\sigma_{\rm min}}\sim \frac{1}{-a\sigma_{\rm min}}\,.
\label{lhside}
\ee
Here we have used the fact that at $R=R_*$, by definition, the derivative of the leading exponential equals the derivative of the rest. For KKLT, this implies
$a(\sigma_*-\sigma_{\rm min})\sim 1$, that is, 
\be
\frac{4}{3}\frac{\left\vert a(R_*-R_{\rm min})\right\vert}{R_*}R_*^{-4/3}\sim 1\,.
\ee
For the limiting case $R_*^{4/3} \sim 10^{-30}$ and with $a\sim 1$, this reduces to
\be
\frac{R_{\rm min}-R_*}{R_*}\sim 10^{-30}\,,
\ee
about 100 times smaller than $(R_{\rm solar\; system}-R_\odot)/R_*\sim 10^{-28}$. 

Substituting~(\ref{lhside}) in~(\ref{rconstra}), we obtain the constraint
\be
|a|\sigma_{\rm min}>10^6\,.
\label{asiglimit}
\ee
We note that this is a much weaker constraint than~(\ref{Rkcond}) on $\gamma R_*^{-k}=|a|\sigma_*$.
We should have expected to obtain again a constraint on the same effective exponent, since for KKLT the leading 
exponential and the minimum are related, being generated by two exponentials of similar form.

Another constraint of interest,~(\ref{rho*}), was obtained by demanding that, for the smallest density of relevance, the average density in the Universe, we are at most at $R_*$, the
limit of the dominance of the leading exponential. This ensured that the field value is not on the steep exponential everywhere in the Universe. In other words, the density $\rho_*$ corresponding to $R_*$ must be greater than the cosmic density, 
\be
\rho_*=R_*\left\vert\frac{{\rm d} V}{{\rm d}R}(R_*)\right\vert\geq H_0^2M_{\rm Pl}^2\,.
\label{rhoconstra}
\ee
Taking the derivative of the KKLT potential at $R_*$, (\ref{approxi}), and dropping terms subleading by $1/(a\sigma_{\rm min})\sim 10^{-14}$, we get
\bea
\nonumber
R_* \frac{{\rm d} V}{{\rm d}R}(R_*) &=& R_*\frac{{\rm d}\sigma}{{\rm d}R}(R_*)\frac{{\rm d}V}{{\rm d}\sigma}(\sigma_*) \\
\nonumber
&\simeq & -\frac{4A^2|a|^3}{9M_{\rm Pl}^2}e^{|a|(\sigma_*+\sigma_{\rm min})}(e^{|a|(\sigma_*-\sigma_{\rm min})}-1)\\
&\simeq & -\frac{4A^2|a|^3}{9M_{\rm Pl}^2} e^{|a|(\sigma_*+\sigma_{\rm min})}.
\label{intermedi}
\eea
From the approximate equation~(\ref{minimumcond}) at the minimum, $W_0\simeq 2A|a|\sigma_{\rm min}e^{|a|\sigma_{\rm min}}/3$, and~(\ref{minimumpot}), we obtain 
\be
A^2e^{|a|(\sigma_*+\sigma_{\rm min})}\sim A^2e^{2|a|\sigma_{\rm min}}\simeq \frac{9}{4}\frac{W_0^2}{a^2\sigma_{\rm min}^2}\simeq \frac{6M_{\rm Pl}^2\sigma_{\rm min}}{a^2}|V_0|\,,
\ee
where we have used $|a|(\sigma_*-\sigma_{\rm min})\sim 1$. Substituting into~(\ref{intermedi}), we obtain
\be
R_* \left\vert\frac{{\rm d} V}{{\rm d}R}(R_*)\right\vert \sim |V_0||a|\sigma_{\rm min}\,.
\ee
The constraint~(\ref{rhoconstra}) hence translates to
\be
|V_0| \;\gsim\; \frac{10^{-120}}{|a|\sigma_{\rm min}}M_{\rm Pl}^4\,.
\label{v0constra}
\ee
Unfortunately, because the bound~(\ref{asiglimit}) on $-a\sigma_{\rm min}$ works in the opposite way, we cannot derive a constraint on $V_0$ itself.
At best, replacing the limiting value of $|a|\sigma_{\rm min} \sim 10^{30}$ implied by~(\ref{Rkcond}), we obtain
\be
|V_0| \;\gsim\; 10^{-150} M_{\rm Pl}^4\,.
\ee
As in the previous subsection, we can solve for $W_0$ using~(\ref{minimumpot}), assuming the same limiting value $|a|\sigma_{\rm min} \sim 10^{30}$:
\be
W_0\;\gsim\;  10^{-30}M_{\rm Pl}^3\,.
\ee
Similarly, writing $Ae^{-a\sigma} = M_{\rm Pl}^3e^{|a|(\sigma-\sigma_0)}$ as in~(\ref{sigma0}), we have $\sigma_0\sim -\log A/M_{\rm Pl}^3\sim 10^{30}$.

Meanwhile, as in Sec.~\ref{consastro} we can use~(\ref{v0constra}) to derive a constraint on mass around the minimum of the potential,
\be
m_{\rm cosmo}^2\simeq \frac{g^2R_*^2}{M_{\rm Pl}^2} \left.\frac{{\rm d}^2V}{{\rm d}R^2}\right\vert_{R = R_{\rm min}}\,,
\label{mmin}
\ee
but this time without assuming the quadratic form. From~(\ref{sigmamass}), we obtain
\be
\left.\frac{{\rm d}^2V}{{\rm d}R^2}\right\vert_{R = R_{\rm min}} \simeq \left(\frac{{\rm d}\sigma}{{\rm d}R}(R_{\rm min})\right)^2\frac{{\rm d}^2V}{{\rm d}\sigma^2}(\sigma_{\rm min})\simeq \frac{32}{9}R_{\rm min}^{-14/3}a^2|V_0|\,.
\ee
Using $R_*\simeq R_{\rm min}$ and assuming $g\sim {\cal O}(1)$,~(\ref{mmin}) reduces to
\be
m^2_{\rm cosmo} \sim a^2R_{\rm min}^{-8/3} \frac{|V_0|}{M_{\rm Pl}^2} = a^2\sigma_{\rm min}^2 \frac{|V_0|}{M_{\rm Pl}^2} \,.
\ee
Our lower bound on $m_{\rm cosmo}$ readily follows from~(\ref{v0constra}):
\be
m_{\rm cosmo} \;\gsim\;  10^{-60} \sqrt{|a|\sigma_{\rm min}}M_{\rm Pl} \;\gsim\; 10^{15}H_0\,,
\ee
where we have used $|a|\sigma_{\rm min} \;\gsim\; 10^{30}$, which follows from~(\ref{Rkcond}) with $R_*\simeq R_{\rm min}$.

We see that while the constraints on the parameters of the KKLT potential are not modified too dramatically, the constraints on the Compton 
wavelength for cosmological densities is, as we go from a megaparsec scale for the quadratic potential to a scale within the Earth's orbit for KKLT.
Thus in order to get an interesting cosmology we would need to modify the KKLT potential with some extra terms, such that one can have a 
valid quadratic approximation for $R>R_*$.

\section{Conclusions}
\label{conclude}
\ \ \ \ \
We have attempted to find a string compactification whose low energy dynamics includes a scalar that exhibits a chameleon mechanism to hide itself from local experiments.  We found that the KKLT potentials are suitable for this purpose, and that the volume modulus of the compactification can act as a chameleon, given the right choice of parameters.  We then looked at experimental constraints coming from tests of gravity, and used these to put bounds on the KKLT parameters.  We find that there are regions of parameters for which the chameleon is safe from detection in all tests so far, yet still has the potential to produce phenomenology different from that of General Relativity.  
Of course, for a full string theory embedding one would have to find specific models with all the scalars stabilized, and obtain for the KKLT 
superpotential both $a<0$, $a\sim {\cal O}(1)$, and $A\sim e^{-10^{30}}\sim e^{-|a|\sigma_{\rm min}}$, or otherwise another potential fitting within our general class.

This raises the question of how generic the chameleon mechanism is in the string landscape.  Usually the fact that we do not observe fundamental scalars in the lab is explained by making the scalars massive through some kind of stabilization, but the chameleon mechanism teaches us that there are other ways to hide scalars besides giving them a very short range.  This opens up a wider range of phenomenologically viable stabilization scenarios.

Note that our general model has the usual chameleon fine-tuning arising from the constraint (\ref{vaccons}). In the application to the example of the
KKLT potential, the fine-tuning seems stronger, however the $e^{\sigma}\sim e^{10^{30}}$ superpotential factor arises from a KK scale of 
$E\sim 10^{11}GeV$, which is certainly fine-tuned with respect to the Planck scale, but not too much. A prefactor $A$ with $\log A\sim -10^{30}$
then arises naturally from imposing that the minimum of the superpotential (\ref{minimumpot}) is not too too large in absolute value. So while 
our KKLT example is fine-tuned, it is not worse than the fine-tuning of having an intermediate KK scale, together with the usual cosmological constant 
problem (the positive minimum is at $\sim 10^{-122}M_P^4$, and the needed AdS supersymmetric minimum $V_0$ is of a similar value).

{\bf Acknowledgements} We thank Robert Brandenberger, Anne Davis, Edward Doheny, James Halverson and Mark Trodden for helpful discussions.
The work of J.K. and K.H. is supported in part by  funds from the University of Pennsylvania and the Alfred P. Sloan Foundation.

%\newpage

\bibliographystyle{utphys}
\bibliography{sympaper20}

\end{document}